\documentclass{rspublic}
\usepackage{graphicx}

\title[Lattice Boltzmann simulations of contact line motion]{Lattice Boltzmann simulations of contact line motion in a liquid--gas system}
\author[A. J. Briant, P. Papatzacos and J. M. Yeomans]{A. J. Briant, P. Papatzacos\thanks{Present adress: Stavanger University College, Stavanger, Norway} and J. M. Yeomans}
\affiliation{Theoretical Physics, Oxford University, 1 Keble Road, Oxford OX1 3NP, UK}
\label{firstpage}

\begin{document}

\maketitle
\begin{abstract}{Mesoscale modelling, lattice Boltzmann, wetting, droplet dynamics, dynamic contact angle}
We use a lattice Boltzmann algorithm for liquid--gas coexistence to investigate the steady state interface profile of a droplet held between two shearing walls. The algorithm solves the hydrodynamic equations of motion for the system. Partial wetting at the walls is implemented to agree with Cahn theory. This allows us to investigate the processes which lead to the motion of the three-phase contact line. We confirm that the profiles are a function of the capillary number and a finite size analysis shows the emergence of a dynamic contact angle, which can be defined in a region where the interfacial curvature tends to zero.
     
\end{abstract}

\section{Introduction}

When a liquid drop in equilibrium with its vapour is placed in contact with a flat surface the equilibrium shape defines a static contact angle, $\theta_w$, between the liquid--vapour interface and the surface. $\theta_w$ is determined by a balance between the fluid--fluid and solid--fluid interactions. The surface may be wet ($\theta_w=0$), partially wet ($0<\theta_w<\pi$) or dry ($\theta_w=\pi$) (de Gennes 1985). 

If the droplet is pushed over the substrate its steady state profile defines two \emph{dynamic} contact angles (advancing and receding) different, in general, from the static angle. The shape of the moving droplet is difficult to investigate analytically because the classical continuum hydrodynamic equations of motion with the usual no slip condition at the surface predict a singularity in the stress at the contact line (Huh \& Sriven 1971; Dussan \& Davis 1974).

In analytic work to remedy this problem it has been proposed that a slip condition may hold at the three phase point and matched asymptotic expansions have been used to describe the interfacial configuration and flow fields near to the boundary (Hocking 1977). More recently Seppecher (1996) used a diffuse interface model with a no slip condition to provide a mesoscopic description of contact line motion. By matching a diffuse interface solution close to the contact line to a sharp interface solution far from it, Seppecher was able to show how the dynamic contact angle is related to the capillary number and to the static angle.

Numerical modelling of contact line motion has also proved difficult because of the widely differing length scales involved in the problem. Molecular dynamics simulations give useful information on the local boundary conditions but are unable to reach length and time scales on which the dynamic contact angle can be measured. Numerical solutions of the Navier--Stokes equations which asssume a sharp interface again suffer from the problem of an infinite stress at the contact line.

It therefore seems appropriate to investigate the extent to which emergent mesoscale modelling techniques allow modelling of the dynamics of droplet motion. These appproaches incorporate a diffuse interface, thus removing any contact line singularity. They are also able to address longer length and time scales than molecular dynamics, albeit at the expense of a lack of molecular detail. A main aim of this paper is to ask whether it is possible to meaningfully define an advancing and receding dynamic contact angle within the framework of one mesoscale approach, lattice Boltzmann simulation (Chen \& Doolen 1998; Succi 2001).

We use the one-component, non-ideal lattice Boltzmann model proposed by Swift \emph{et al} (1996) to investigate motion of the contact line. The simulations model a droplet held between parallel plates, which impose a shear flow on the system. We have developed boundary conditions which allow us to control the wetting behaviour at the plates. In  particular, the static contact angle agrees with that predicted analytically by Cahn theory (Cahn 1977). The simulations show that, in the steady state, a dynamic contact angle can be defined away from the wall. However, even in two dimensions, rather large lattices are needed to do this. 

This paper is organised as follows. In \S\ref{method} we outline the lattice Boltzmann scheme used, \S\ref{wetting} describes the boundary conditions necessary to simulate wetting and \S\ref{spreading} presents the results of the simulations. Finally in \S\ref{discussion} we discuss the results and draw conclusions.

\section{Method}
\label{method}
Lattice Boltzmann simulations (Chen \& Doolen 1998; Succi 2001; Swift \textit{et al} 1995, 1996) solve the Navier--Stokes equations by following the evolution of a set of distribution functions, $f_{\sigma i}(\bf{x}, t)$, which represent the density at time $t$ and lattice site ${\bf x}$ which is travelling with velocity ${\bf e}_{\sigma i}$. The velocity vectors $\bf{e}_{\sigma i}$ are such that the distribution functions advect to neighbouring lattice sites $\bf x + \Delta {\bf x}$ in the time interval $\Delta t$. The velocity  vector subscripts $\sigma, i$ and $\alpha$ are used to specify a vector's magnitude, direction and Cartesian components. For a square lattice and a nine-speed lattice Boltzmann model which we consider here, $\sigma=0$ for the zero velocity vector, $\sigma=1, i=1, 2 \ldots 4$ for vectors to nearest-neighbour sites and $\sigma=2, i=1, 2 \ldots 4$ for vectors to next-nearest neighbour sites. Physical quantities are related to moments of $f_{\sigma i}$. The density $n$ and fluid velocity ${\bf u}$ are defined by 

\begin{gather}
\sum_{\sigma i} f_{\sigma i} = n, \\
\sum_{\sigma i} f_{\sigma i}e_{\sigma i \alpha} = n u_{\alpha}.
\end{gather}

The distributions $f_{\sigma i}$ are evolved according to a lattice Boltzmann equation assuming a single relaxation time approximation

\begin{equation}\label{LBE}
f_{\sigma i}({\bf x} + {\bf e}_{\sigma i}\Delta t, t + \Delta t)-f_{\sigma i}({\bf x}, t) =-\frac{\Delta t}{\tau}(f_{\sigma i}-f_{\sigma i}^{eq})
\end{equation}
where $\tau$ is the relaxation time and $f^{eq}_{\sigma i}$ is an equilibrium distribution function. The equilibrium distribution determines the physics inherent in the simulation. A power series in the local velocity is assumed (Swift \textit{et al} 1996)
\begin{equation}
f^{eq}_{\sigma i} = A_{\sigma} + B_{\sigma} e_{\sigma i \alpha} u_{\alpha} + C_{\sigma} u^2 + D_{\sigma}e_{\sigma i \alpha}e_{\sigma i \beta}u_{\alpha}u_{\beta} + G_{\sigma \alpha \beta} e_{\sigma i \alpha}e_{\sigma i \beta}
\end{equation}
where summation over repeated Cartesian indices is understood.

The coefficients $A_\sigma$, $B_\sigma$, $C_\sigma$, $D_\sigma$ and $G_{\sigma \alpha \beta}$ are determined by placing constraints on the moments of $f_{\sigma i}^{eq}$. In order that the collision term in equation (\ref{LBE}) conserves mass and momentum the first two moments of $f_{\sigma i}^{eq}$ are constrained by 

\begin{gather}
\label{sumf}\sum_{\sigma i} f^{eq}_{\sigma i} = n,\\
\label{sumfe}\sum_{\sigma i} f^{eq}_{\sigma i}e_{\sigma i \alpha} = n u_{\alpha}.
\end{gather}

The next moment of $f_{\sigma i}^{eq}$ is chosen such that the continuum macroscopic equations approximated by the evolution scheme (\ref{LBE}) correctly describe the hydrodynamics of a one-component, non-ideal fluid. This gives
\begin{equation}
\label{sumfee}\sum_{\sigma i} f^{eq}_{\sigma i}e_{\sigma i \alpha} e_{\sigma i \beta}= P_{\alpha \beta} + n u_{\alpha} u_{\beta} + \nu[u_{\alpha}\partial_{\beta}(n) + u_{\beta} \partial_{\alpha}(n) + u_{\gamma}\partial_{\gamma}(n)\delta_{\alpha \beta}]
\end{equation}
where  $\nu=(\tau-\Delta t/2)/3$  is the kinematic shear viscosity and $P_{\alpha \beta}$ is the pressure tensor. The first formulation of the model omitted the third term in equation (\ref{sumfee}) and was not Galilean invariant. Holdych \textit{et al} (1998) showed that the addition of this term led to any non-Galilean invariant terms being of the same order as finite lattice corrections to the Navier--Stokes equations. In order to fully constrain the coefficients $A_\sigma$, $B_\sigma$, $C_\sigma$, $D_\sigma$ and $G_{\sigma \alpha \beta}$ a fourth condition is needed (Huo \textit{et al} 1995)
\begin{equation}
\label{sumfeee}\sum_{\sigma i} f^{eq}_{\sigma i}e_{\sigma i \alpha} e_{\sigma i \beta}e_{\sigma i \gamma}= \frac{n}{3}(u_{\alpha}\delta_{\beta \gamma}+u_{\beta}\delta_{\alpha \gamma}+u_{\gamma}\delta_{\alpha \beta}).
\end{equation}
The analysis of Holdych \textit{et al} (1998) shows that the evolution scheme (\ref{LBE}) approximates the following Navier--Stokes level equation:
\begin{multline}
\label{navierstokes}
\partial_t (n u_\alpha) + \partial_\beta (n u_\alpha u_\beta) = - \partial_\beta P_{\alpha \beta} + \nu  \partial_\beta \bigl[ n (\partial_\beta(u_\alpha) + \partial_\alpha(u_\beta) + \partial_\gamma (u_\gamma) \delta_{\alpha \beta})\bigr] \\
- 3\nu \partial_\beta \bigl[ u_\alpha \partial_\gamma P_{\beta \gamma} + u_\beta \partial_\gamma P_{\alpha \gamma} + \partial_\gamma (n u_\alpha u_\beta u_\gamma) \bigr] \\
- 3 \nu \partial_\beta \bigr[(\partial_n P_{\alpha \beta})\partial_\gamma(n u_\gamma)\bigr]\\
-3\nu^2 \partial_\beta \bigl[u_\alpha\partial_\gamma\bigl(u_\beta\partial_\gamma (n) + u_\gamma \partial_\beta (n) + u_\lambda \partial_\lambda (n) \delta_{\alpha \beta}\bigr)\bigr]\\
-3\nu^2 \partial_\beta \bigl[u_\beta\partial_\gamma\bigl(u_\alpha\partial_\gamma (n) + u_\gamma \partial_\alpha (n) + u_\lambda \partial_\lambda (n) \delta_{\alpha \beta}\bigr)\bigr]\\
+3\nu^2 \partial_\beta \bigl[\partial_t\bigl(u_\alpha\partial_\beta (n) + u_\beta \partial_\alpha (n) + u_\lambda \partial_\lambda (n) \delta_{\alpha \beta}\bigr)\bigr].
\end{multline}

The thermodynamics of the fluid enter the lattice Boltzmann simulation via the pressure tensor $P_{\alpha \beta}$ (Rowlinson \& Widom 1982). For a system without surfaces the equilibrium properties of the fluid can be described by a Landau free energy functional of the form (Landau \& Lifshitz 1958)
\begin{equation} 
\label{bulkfreeenergy}
\Psi_b = \int \rd V \bigl[ \psi(T, n) + \frac {\kappa}{2} (\partial_{\alpha}n)^2 \bigr] 
\end{equation}
where $\kappa$ is related to the surface tension, and (Rowlinson \& Widom 1982)
\begin{equation} 
\label{Wdefn}
\psi(T, n) = W(T, n)+\mu_b(T) n - p_b(T). 
\end{equation}
Here, $\mu_b$ and $p_b$ are the chemical potential and pressure in the bulk, and W is a non-negative function of $n$ that vanishes, along with $\partial W/\partial n$, when $n$ is equal to the liquid bulk density $n_l$ or to the gas bulk density $n_g$. It then follows that (Rowlinson \& Widom 1982) 
\begin{equation}
P_{\alpha \beta}({\bf x}) = \delta_{\alpha \beta} p({\bf x}) + \kappa (\partial_{\alpha}n)(\partial_{\beta}n) 
\end{equation}
with
\begin{equation}
p({\bf x}) = p_0 - \kappa n \partial_{\gamma \gamma}n - \frac{\kappa}{2}(\partial_\gamma n)^2
\end{equation}
where $p_0 = n \partial_n \psi(T, n) - \psi(T, n)$ is the equation of state of the fluid. However, to incorporate wetting the choice of free energy must be generalised to include surface terms and we now consider this case.

\section{Wetting}
\label{wetting}
When a liquid--gas interface meets a solid wall the angle, $\theta_w$, between the interface and the wall, measured in the liquid, is determined by the liquid--gas, solid-liquid and solid-gas surface tensions, $\sigma$, $\sigma_{sl}$ and $\sigma_{sg}$ according to Young's equation (Young 1805)
\begin{equation}\label{young}
\cos \theta_w = \frac{\sigma_{sg} - \sigma_{sl}}{\sigma}.
\end{equation}
In this section our aim is to define lattice Boltzmann boundary conditions which reproduce Young's equation in equilibrium. The solid--gas and solid--liquid surface tensions will be related to an additional term in the Landau free energy functional which describes the interactions at the surface between the solid and the fluid. To this end we follow Cahn (1977) and introduce an additional surface term into the free energy
\begin{equation} 
\label{wallfreeenergy}
\Psi_s = \int \rd S\;\;\Phi(n_s) 
\end{equation}
where $n_s$ is the fluid density at the wall. Following de Gennes (1985) we expand $\Phi(n_s)$ as a power series in $n_s$. In addition, and following Seppecher (1996), we keep only the first order term and write $\Phi = -\phi_1 n_s$ since this turns out to be sufficient for the partial wetting scenarios that we will want to consider.

We now summarise how the wetting angle $\theta_w$ can be written in terms of the wetting potential $\phi_1$ (Cahn 1977; Papatzacos 2001). To do this we calculate the surface tensions by considering a one dimensional problem where one phase of a non-ideal fluid occupies the region $x>0$ with a solid wall at $x=0$. Far from the wall the fluid density will be $n_b$ (where $n_b$ is the bulk liquid density $n_l$ or the bulk gas density $n_g$), while at the wall $n=n_s$ with $n_s$ undetermined as yet. In one dimension the free energy functional (\ref{bulkfreeenergy}) together with (\ref{wallfreeenergy}) reduces to 
\begin{equation}
\label{fullfreeenergy}
\Psi_b + \Psi_s = \int \rd x \,\bigl[ \psi(T, n) + \frac {\kappa}{2} (\partial_x n)^2 \bigr] - \phi_1 n_s.
\end{equation}

Minimising (\ref{fullfreeenergy}) subject to natural boundary conditions leads to two conditions
\begin{gather}
\label{EL}\frac{\partial \psi}{\partial n} - \mu_b - \kappa \frac{\rd^2 n}{\rd x^2} = 0 \qquad \text{for $x>0$},\\ 
\label{BC}\kappa \bigl(\frac{\rd n}{\rd x} \bigr) = \frac{\rd\Phi(n_s)}{\rd n_s} = -\phi_1 \qquad \text{at $x=0$}.
\end{gather}
Equation (\ref{EL}) is the usual Euler--Lagrange equation and (\ref{BC}) is a boundary condition valid at $x=0$. A first integral for equation (\ref{EL}) is 
\begin{equation}
\label{first}\frac{\kappa}{2}\bigl(\frac{\rd n}{\rd x} \bigr)^2 = \psi - \mu_b n + p_b = W(n)
\end{equation}
suppressing for now the $T$ dependence of $W$. We can determine $n_s$ by substituting (\ref{first}) into (\ref{BC}) giving
\begin{equation}
\label{nscondition}-\phi_1 = \pm \sqrt{2 \kappa W(n_s)}.
\end{equation}

Consider first $\phi_1>0$. Equation (\ref{nscondition}) has four solutions ($n_1<n_2<n_3<n_4$) if $\phi_1$ is smaller than the height of the double well function defined by $\sqrt{2 \kappa W}$ (Cahn 1977). The value of $n_s$ is obtained from one of these four solutions as the one which minimises the solid--fluid surface tension
\begin{equation}\label{sftension}
\sigma_{sf} =  - \phi_1 n_s + \int \sqrt{2\kappa W}\;\; \rd n .
\end{equation}
For $\phi_1$ small enough it can be shown that the minimising solutions are $n_2$ if $n_b = n_g$ and $n_4$ if $n_b = n_l$. This gives expressions for the suface tensions
\begin{gather}
\sigma_{sg} = -\phi_1 n_2 + \int_{n_g}^{n_2}\sqrt{2\kappa W}\;\; \rd n,\\
\sigma_{sl} = -\phi_1 n_4 + \int_{n_l}^{n_4}\sqrt{2\kappa W}\;\;  \rd n.
\end{gather}
Similarly, if $\phi_1 < 0$ the minimising solutions are $n_1$ if $n_b = n_g$ and $n_3$ if $n_b = n_l$ and the solid--fluid surface tensions are
\begin{gather}
\sigma_{sg} = -\phi_1 n_1 + \int_{n_1}^{n_g}\sqrt{2\kappa W}\;\; \rd n,\\
\sigma_{sl} = -\phi_1 n_3 + \int_{n_3}^{n_l}\sqrt{2\kappa W}\;\;  \rd n.
\end{gather}
The liquid--gas surface tension $\sigma$ follows from (\ref{fullfreeenergy}) as
\begin{equation}\label{fftension}
\sigma = \int_{n_g}^{n_l}\sqrt{2\kappa W}\;\; \rd n.
\end{equation}

To find a closed form for $\theta_w$ we now choose the excess free energy function $W(n, T)$ to be
\begin{equation}
W(\nu) = p_c (\nu^2 - \beta \tau)^2
\end{equation}
where $\nu = (n-n_c)/n_c$ and $\tau=(T_c-T)/T_c$. $T_c$, $p_c$ and $n_c$ are the critical temperature, pressure and density respectively, and $\beta$ is a constant. With this form for $W(n, T)$ Young's law (\ref{young}) reduces to 
\begin{equation}\label{young2}
\cos \theta_w = \frac{(1+\Omega)^{3/2}-(1-\Omega)^{3/2}}{2}
\end{equation}
where $\Omega  \equiv \phi_1/\beta\tau\sqrt{2\kappa p_c}$.
 
In order to implement this scheme in a lattice Boltzmann simulation equation (\ref{BC}) is imposed on lattice sites which represent the wall. Our approach in similar to that recently taken by Desplat \textit{et al} (2001) in introducing wetting boundary conditions for a binary mixture. Since (\ref{BC}) is an equilibrium condition, it is appropriate to impose it through the equilibrium distribution function, $f^{eq}_{\sigma i}$. The coefficients of $f^{eq}_{\sigma i}$ depend on the local values of $n$, $\bf{\nabla n}$ and $\nabla^2n$ which, in the bulk, are calculated using standard finite difference methods. For a wall parallel to a lattice direction, it is the perpendicular components of $\bf{\nabla n}$ and $\nabla^2n$ which must be  calculated using equation (\ref{BC}). For $\partial_x n$ (where $x$ is the perpendicular direction to the wall) we use the calculated value of  $\phi_1/\kappa$. For $\partial^2_x n$, we use the standard right-handed finite difference formula
 \begin{equation}
 \partial^2_x n|_{i=0} \approx \frac{-3 n_0'+4n_1'-n_2'}{2}
 \end{equation}
 where $n_i'$ is $\partial_x n$ at lattice site $i$. In this formula we substitute for $n_0'$ using equation (\ref{BC}) and calculate $n_1'$ using a standard centred finite difference formula. Finally we have found empirically that the best choice for $n_2'$ is a left-handed finite difference formula taken \textit{back} into the wall
 \begin{equation}
 n_2' \approx \frac{3 n_2-4n_1+n_0}{2}.
 \end{equation}

\begin{figure}
\includegraphics[width=4.5in]{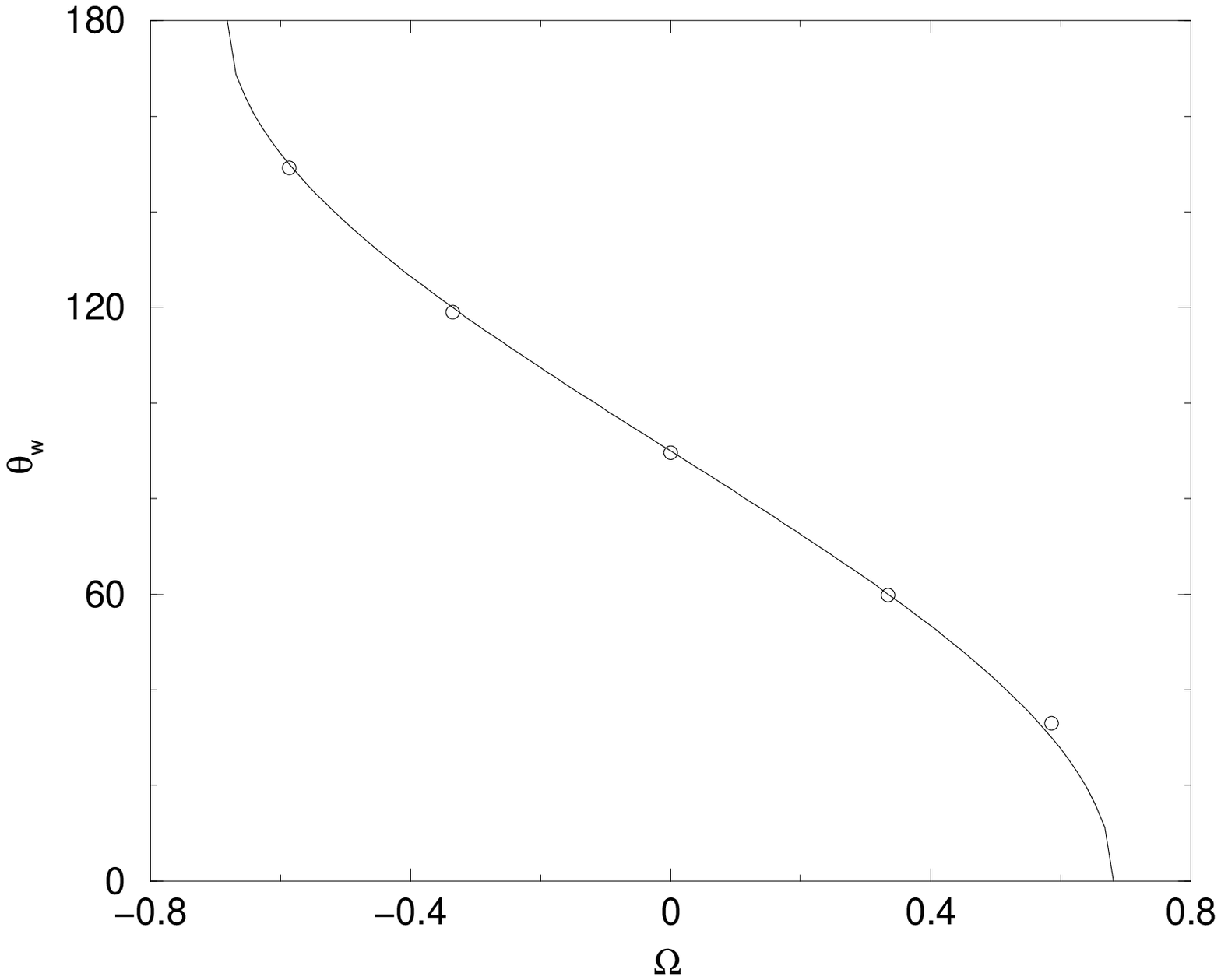}
\caption{Static wetting angle $\theta_w$ plotted as a function of dimensionless wetting potential $\Omega$. Curve: theoretical relation, equation (\ref{young2}); circles: simulation results.}
\label{thetavomega}
\end{figure}

Using this scheme to evaluate $f^{eq}_{\sigma i}$ at wall sites it is possible to control the wetting angle at any flat wall. To validate the method we have simulated a droplet of liquid in equilibrium with its gas on a solid surface. Figure \ref{thetavomega} shows how the observed contact angle varies with $\Omega$. The agreement between simulation and theory is good and therefore we take this method as a basis for simulations of spreading.

\section{Contact line dynamics}
\label{spreading}
We now present the results of simulations exploring the motion of a contact line. Our aim is to show how a dynamic contact angle can be measured using the lattice Boltzmann method.

In his analytic work on contact line motion, Seppecher (1996) considered a droplet on a surface and defined three regions of flow for his matched solution. In the inner zone, a semicircle of radius $R_1$, Seppecher used the Cahn--Hilliard equation to solve for the flow field with the condition that the interface intersected the wall at the static contact angle, $\theta_w$. Interfacial curvature is concentrated in this region and is the cause of mass transfer across the interface, which enables the interface (but not the fluid) to slip relative to the surface. Seppecher matched this inner solution to a solution of the Stokes equation in the intermediate zone. In the intermediate zone, an annulus  of outer radius $R_2 (\gg R_1)$, the interface was considered to be flat and have zero width. The angle this part of the interface makes with the surface is taken to be the apparent dynamic contact angle, $\Phi$---the angle observable in experiments. The intermediate zone is matched to the outer zone which is the region far from the contact line. In the external zone the curvature radius of the interface is much larger than $R_2$ and the external forces on the droplet and geometry of the whole domain determine the flow.

Using diffuse interface simulations, and a droplet of a computationally feasible size, the intermediate zone, which defines the dynamic contact angle, is masked by the curvature of the droplet. Therefore in our simulations we have employed a different geometry which allows us to probe the intermediate zone within the limits of the computational resources available.

\begin{figure}
\begin{center}
\includegraphics{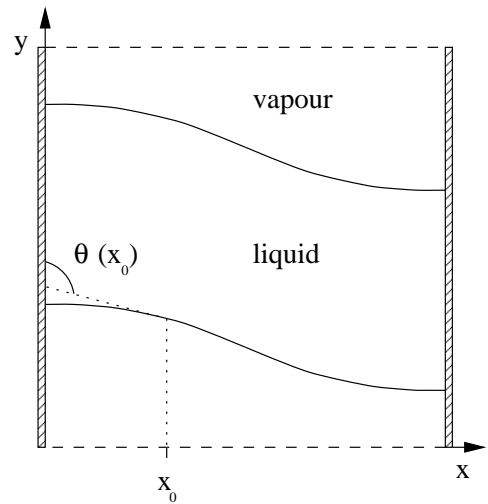}
\end{center}
\caption{Sheared interface profiles showing the definition of $\theta(x_0)$.}
\label{droplet}
\end{figure}

Our simulations model a two dimensional droplet held between two partially wetting walls at $x=0$ and $x=L_x-1$, where $L_x$ is the number of lattice  points in the $x$ direction. The wetting properties of the two walls may be varied independently. The system is $L_y$ lattice sites in the $y$ direction where periodic boundary conditions close the system. The droplet is allowed to reach equilibrium with its gas, and then a shear flow is applied, by imposing velocities $\pm V_0$ on the walls. The system achieves a steady state and the interface profile is recorded by measuring the angle, $\theta (x)$, the interface makes with the wall at $x=0$ (see figure \ref{droplet}).

\begin{figure}
\includegraphics[width=4.5in]{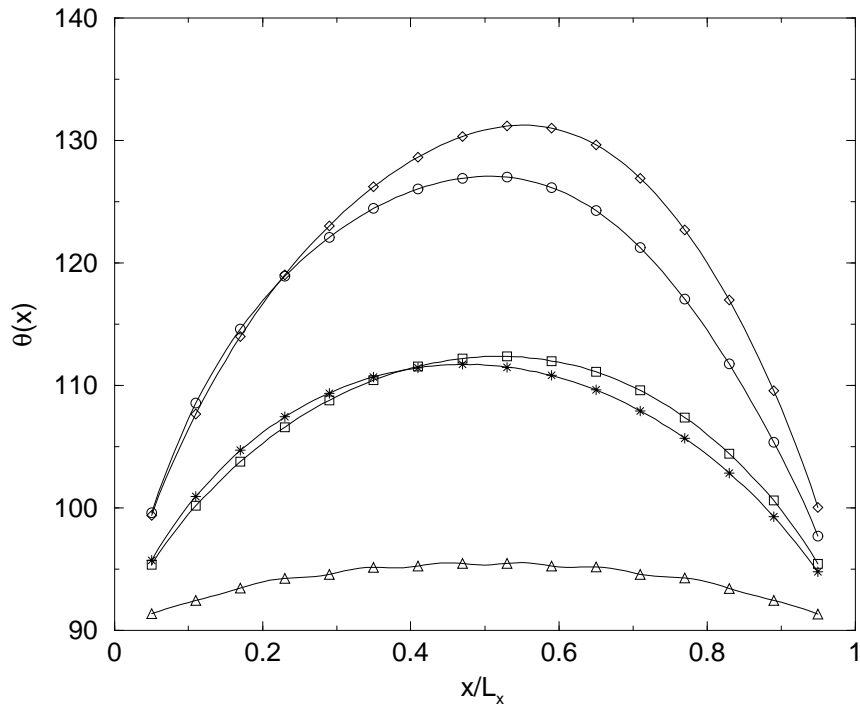}
\caption{Interface profiles $\theta (x)$ plotted against  $x/L_x$ for a reference capillary number, $Ca=0.085$ ($\triangle$), and capillary numbers four (case A:$\ast$; case B:$\square$) and seven (case A:$\circ$; case B:$\diamond$) times greater.}
\label{cascaling}
\end{figure}

Initial simulations focussed on confirming the dependence of the interface profile on the capillary number, $Ca=\eta U/\sigma$ where $\eta$ is the viscosity, $U$ is the speed of the interface relative to the wall and $\sigma$ is the surface tension. For small systems ($L_x = L_y = 100$) we varied the capillary number from a reference value by independently varying the viscosity (case A) or the shear rate (case B): The aim was to show that $\theta(x)$ depends only on the product $\eta V_0$. Figure \ref{cascaling} shows steady state interface profiles for a reference value of the capillary number, $Ca=0.085$, and for capillary numbers four and seven times greater than this value. This figure reveals that $\theta(x)$ scales approximately with capillary number, although there are some surprising discrepancies. However, the discrepancies between cases A and B for each capillary number can be explained as being due to spurious velocities in the lattice Boltzmann model (which are a lattice artifact). The case of increasing $\tau$ (case A) converges on the results of $\tau=0.8$ (case B) if the resolution of the grid is increased (Briant, unpublished data). This value of $\tau$ minimises the spurious velocities due to cancellation of higher order terms in the Chapman-Enskog expansion, as noted by Swift \textit{et al} (1996).

\begin{figure}
\includegraphics[width=4.5in]{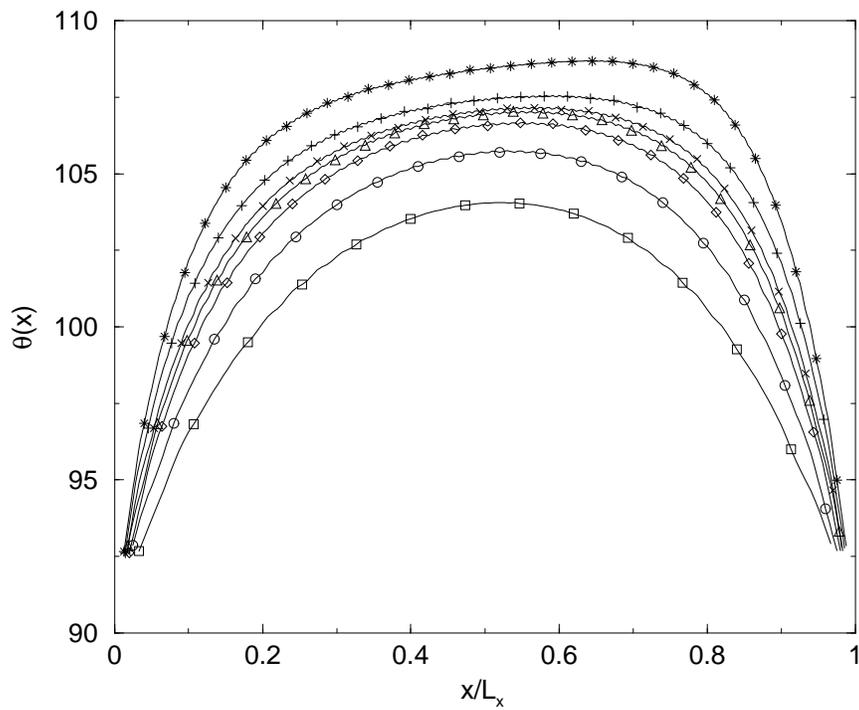}
\caption{Interface profiles $\theta (x)$ plotted against $x/L_x$ for fixed capillary number. $L_x$ values are $150$ ($\square$), $200$ ($\circ$), $250$ ($\diamond$), $275$ ($\triangle$), $300$ ($\times$), $350$ ($+$) and $400$ ($\ast$).}
\label{9090finite}
\end{figure}

We now address how to measure a dynamic angle. From Seppecher (1996), we expect an intermediate zone, where the curvature of the interface is zero, to develop far from the contact line. Therefore we perform the shearing experiments described above and vary the number of lattice sites in the $x$ direction, to see whether such an intermediate zone can be observed. Figure \ref{9090finite} shows the interface profiles achieved for two neutrally wetting walls ($\theta_w=90^\circ$) and lattice sizes of $L_x=150, 200, 250, 275, 300, 350$ and $400$.

\begin{figure}
\includegraphics[width=4.5in]{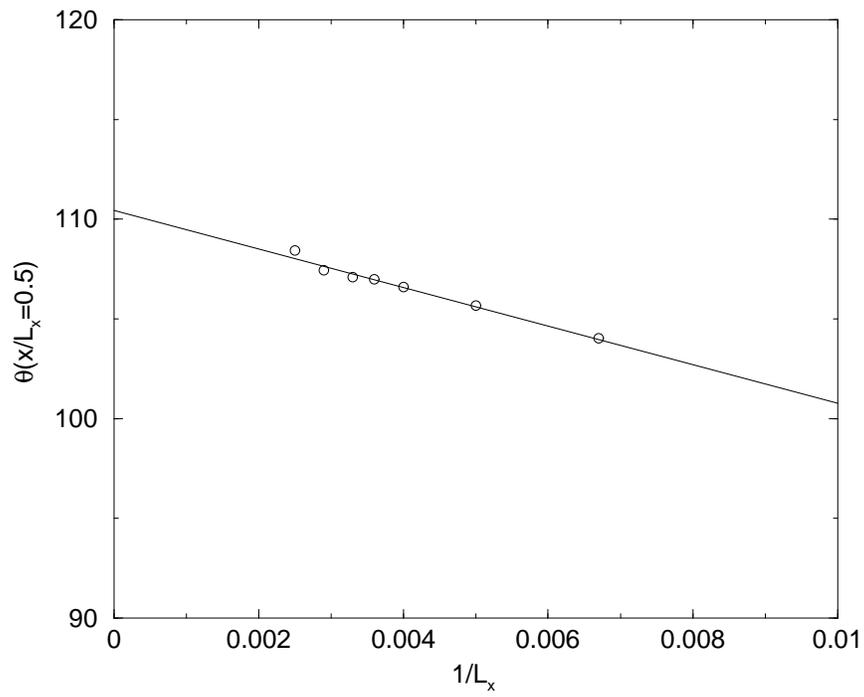}
\caption{Mid point interface angles $\theta(x/L_x=0.5)$ plotted against $1/L_x$.}
\label{invLplot}
\end{figure}

Figure \ref{9090finite} shows that as the distance between the walls is increased, the angle in the central region of the domain approaches a limiting value. This is more clearly indicated in figure \ref{invLplot} where we have plotted $\theta(x/L_x=0.5)$ against $L_x^{-1}$ for each set of data. Figure \ref{invLplot} shows that the midpoint angle is converging to a well-defined value for an infinite system. Identifying the extrapolated value as the dynamic angle gives $\Phi=110^\circ$.

\section{Discussion}
\label{discussion}

In this paper we have described a lattice Boltzmann scheme for the simulation of partial wetting and of moving contact lines in a liquid gas system. Wetting boundary conditions have been defined which enable the contact angle of the interface to be controlled in a way consistent with Cahn theory. We have used results for the shape of a droplet held between sheared parallel plates, together with a finite size analysis, to investigate the dynamic contact angle. The approach demonstrates that a lattice Boltzmann model with hydrodynamic no slip boundary conditions can lead to a well defined dynamic contact angle. However, the limitation of a diffuse interface approach is that it is difficult to model large domains because of the magnitude of the computational demand. Therefore measuring a meaningful dynamic contact angle for, say, a droplet moving along a surface would prove very difficult.

Future directions for the work described in this paper include the study of different wetting conditions at the sheared walls and finding the functional relationship between the capillary number and the dynamic contact angle. It is of interest that recent molecular dynamics simulations of binary fluids (Denniston \& Robbins 2001) have questioned the validity of no slip boundary conditions. Although mesoscale simulations are unable to investigate the molecular basis for imposing a given boundary condition they provide an ideal tool for understanding how a change in the boundary conditions employed will affect the interface profile.

Jones \textit{et al} (1999) have used a different mesoscale approach,  dissipative particle dynamics, to investigate how a droplet in a binary fluid is pulled from a wall under shear flow.  Dissipative particle dynamics simulations differ from lattice Boltzmann in that they include fluctuations. It would be interesting to compare results from the two approaches for the geometry described here. Recently Fan \textit{et al} (2001) have used the two component lattice Boltzmann method developed by Shan \& Chen (1993) to study the contact line dynamics of a pressure driven binary fluid in a capillary tube. They found that the cosine of the contact  angle varies linearly with the speed of the contact point for two different wetting conditions. Other lattice Boltzmann investigations of contact line motion for binary fluids have been given by Grubert \& Yeomans (1999) and Desplat \textit{et al} (2001) using a free energy approach similar to that given here. These authors point out that for binary systems  interspecies diffusion is important in facilitating the motion of the interface near the wall.

\begin{acknowledgements}
We thank A. Balazs, C. Denniston, I. Pagonabarraga and P. Warren for helpful discussions. AB acknowledges EPSRC research studentship 99315535 and a CASE award from Unilever plc.
\end{acknowledgements}

\end{document}